\documentclass[aip, jcp,amsmath,amssymb,reprint,author-numerical]{revtex4-1}
\usepackage{epsfig}
\usepackage{graphicx}
\usepackage{dcolumn}
\usepackage{bm}
\begin{document}

\title[C2-NEB]{Nudged-elastic band method with two climbing images: 
finding transition states in complex energy landscapes}
\thanks{C2-NEB code is available via email.}

\author{Nikolai A. Zarkevich}
\email{zarkev@ameslab.gov}
\affiliation{Ames Laboratory, U.S. Department of Energy, Ames, Iowa 50011-3020}
\author{Duane D. Johnson} 
\affiliation{Ames Laboratory, U.S. Department of Energy, Ames, Iowa 50011-3020}
\affiliation{Materials Science \& Engineering, Iowa State University, Ames, Iowa 50011-2300} 

\date{\today}

\begin{abstract}
The nudged-elastic band (NEB) method is modified with concomitant two climbing images (C2-NEB) to find a transition state (TS) in complex energy landscapes, such as those with serpentine minimal energy path (MEP). 
If a single climbing image (C1-NEB) successfully finds the TS, C2-NEB finds it with higher stability and accuracy. 
However, C2-NEB is suitable for more complex cases, where C1-NEB misses the TS because the MEP and NEB directions near the saddle point are different. 
Generally, C2-NEB not only finds the TS but guarantees that the climbing images approach it from the opposite sides along the MEP, 
and it estimates accuracy from the three images: the highest-energy one and its climbing neighbors. 
C2-NEB is suitable for fixed-cell NEB and the generalized solid-state NEB (SS-NEB). 
\end{abstract}


\maketitle

\section{Introduction}
{\par } The Nudge-Elastic Band (NEB) method has become a workhorse in determining the transition states (TS) and minimum-energy pathways (MEP) 
for both fixed-cell \cite{C1NEB} and generalized solid-state transformations \cite{GSSNEB}, 
especially when implemented within density-functional theory (DFT). 
The available climbing-image algorithm, known as C1-NEB, is a modification of the NEB method,  \cite{NEB,NEBbook} where at each step the highest-energy image climbs. 
Here, to broaden the NEB applicability, we provide an extension to two climbing images (C2-NEB) method that is more reliable and accurate for complex potential-energy landscapes. 

\begin{figure}[t]
\includegraphics{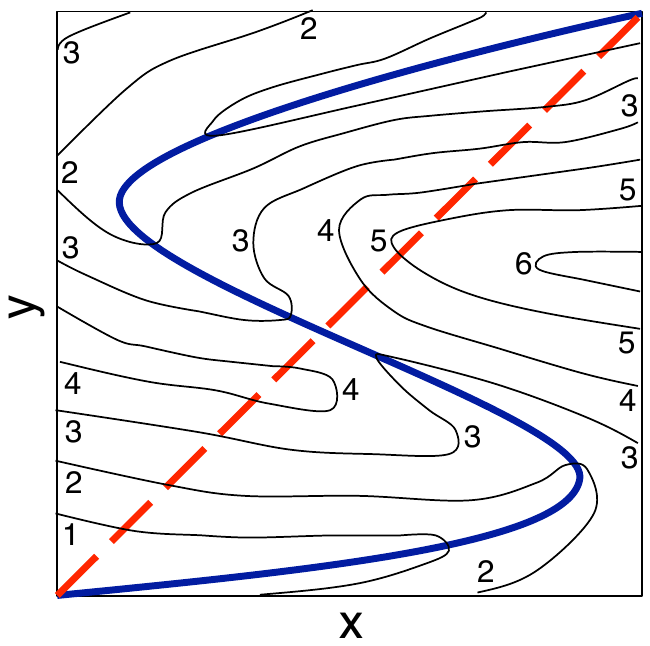}
\caption{\label{Fig1} 
Example of a serpentine MEP (solid blue curve) and a sparse-image NEB direction (dashed red line) near the TS saddle point. A schematic is shown for energy isosurfaces for a system with two degrees of freedom $(x,y)$, with arbitrary relative units. 
}
\end{figure}

\begin{figure}[ht]
\includegraphics{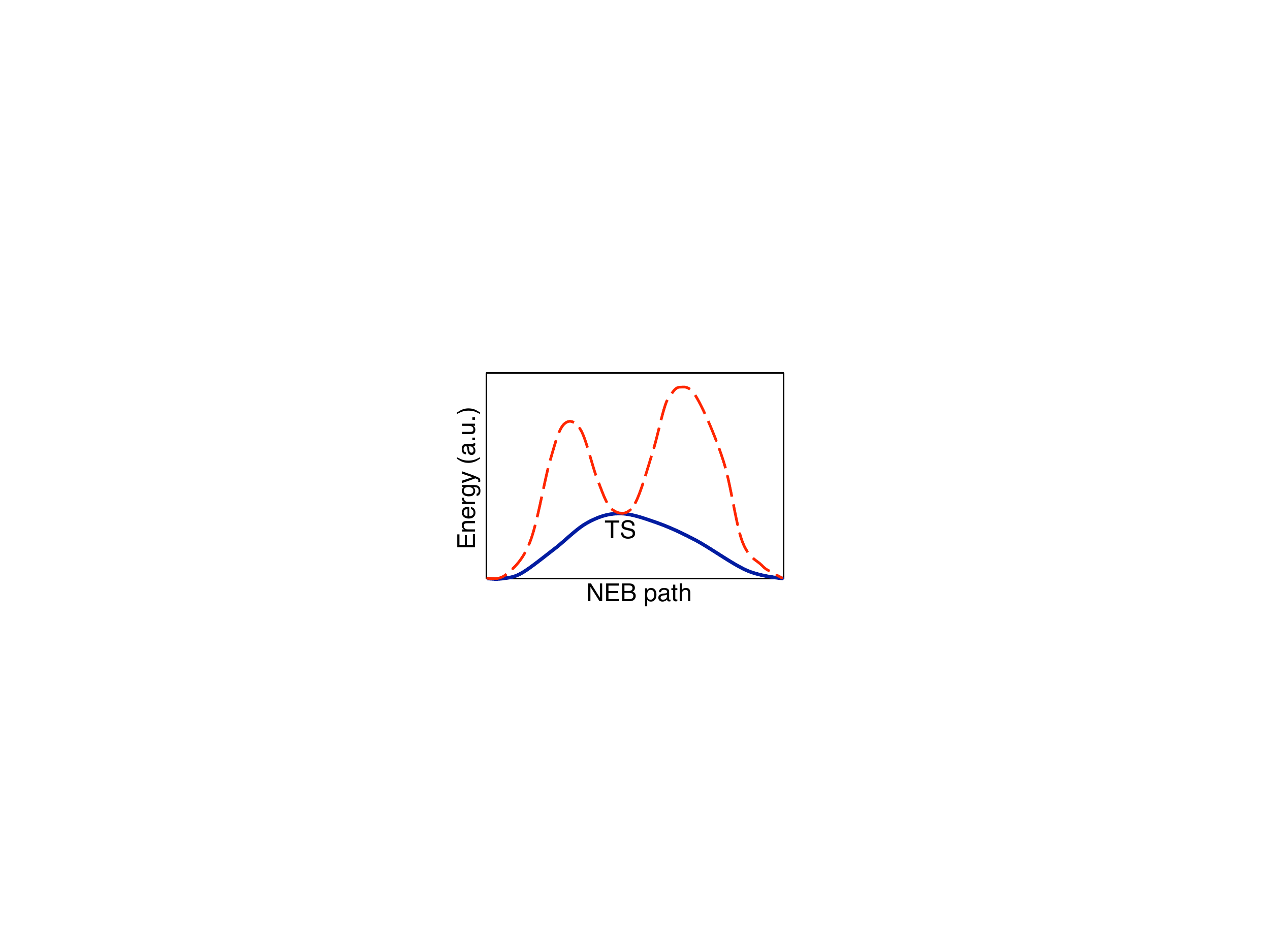}
\caption{\label{Fig2} 
Energy (in arbitrary units) along the NEB path near TS in a C2-NEB (solid blue) and in a sparse-image C1-NEB (dashed red) algorithms. 
Due to different MEP and sparse-image NEB directions (Fig.~\ref{Fig1}), the saddle point is the energy maximum along the path in C2-NEB, but not in a sparse-image C1-NEB. 
}
\end{figure}

{\par } In C2-NEB with two climbing images, the two neighbors of the highest-energy image (HEI) both climb towards TS, while the highest-energy image is nudged to them.
Hence, all three images move towards the TS saddle point, and the two climbing images are guaranteed to approach it along the MEP from opposite directions; their motion along the MEP provides higher stability compared to C1-NEB, see Fig.~\ref{Fig2}, 
with an accurate estimate of the TS energy from energies of the three images.
For a serpentine MEP (Fig.~\ref{Fig1}), C1-NEB can miss the TS, 
while C2-NEB does not require as many images as C1-NEB to find the correct result (Fig.~\ref{Fig2}), and naturally provides a denser grid of images near the saddle point, so it is more efficient and converges faster.

{\par }	To showcase the method, we apply our C2-NEB code implemented within VASP \cite{VASP} 
to study deformation of the shape-memory alloy NiTi, 
in particular its ground-state body-centered orthorhombic (BCO) structure. \cite{BCO} 
As an illustration, we consider two cases, covering success (Fig.~3) and failure (Fig.~4) of the available C1-NEB algorithm. 
In the case of C1-NEB success, C2-NEB finds the same TS. 
In the case of C1-NEB failure, C2-NEB successfully finds the TS, which was missed by C1-NEB (Figs. 1 and 2). 
We conclude that our C2-NEB method is more stable and reliable, with a broader applicability.


\section{Method}
{\par} Our code is implemented as follows. At each molecular dynamics (MD) step, the highest-energy image is found. 
If this HEI is one of the two terminal points, then there is no climbing. If it is next to the fixed terminal point, then C1-NEB is used for this step. Finally, if the HEI does not coincide with the fixed ends or their direct neighbors, then C2-NEB algorithm is used for this step. Because the current highest-energy image is determined at each MD step, the climbing images (and even the algorithm) can change from step to step. However, these changes do not affect the final converged result. 

{\par}Formally, let $N$ moving images be enumerated from 1 to $N$, two fixed terminal images having indices $0$ and $(N+1)$, and at each MD step the index $M$ of the maximal energy image is found. In these notations, our algorithm can be presented by the following pseudo-code. 
{\par}
At each MD step, the algorithm for this step is chosen:
\begin{itemize}
\item     if $1<M<N$, then C2-NEB;
\item    else if $M=1$ or $M=N$, then {C1-NEB};
\item    else NEB without climbing.
\end{itemize}
Each image $I$ decides, if it is climbing at this MD step. In C2-NEB,
\begin{itemize}
\item   if $I=(M-1)$ or $I=(M+1)$, then climbing;
\item   else nudged (not climbing).
\end{itemize}
If C1-NEB is the algorithm, then 
\begin{itemize}
\item if $I=M$, then climbing;
\item  else nudged.
\end{itemize}
In a NEB without climbing, every image is nudged.

{\par}
Our C2-NEB code is available as a replacement of file {\tt{neb.F}} in the NEB code. \cite{C1NEB}   
We combine C2-NEB with DFT implemented in VASP. \cite{VASP}  
The two climbing images method is turned on by setting LCLIMB=.TRUE. in the INCAR file. \cite{VASP}

\begin{figure}
\includegraphics[width=80mm]{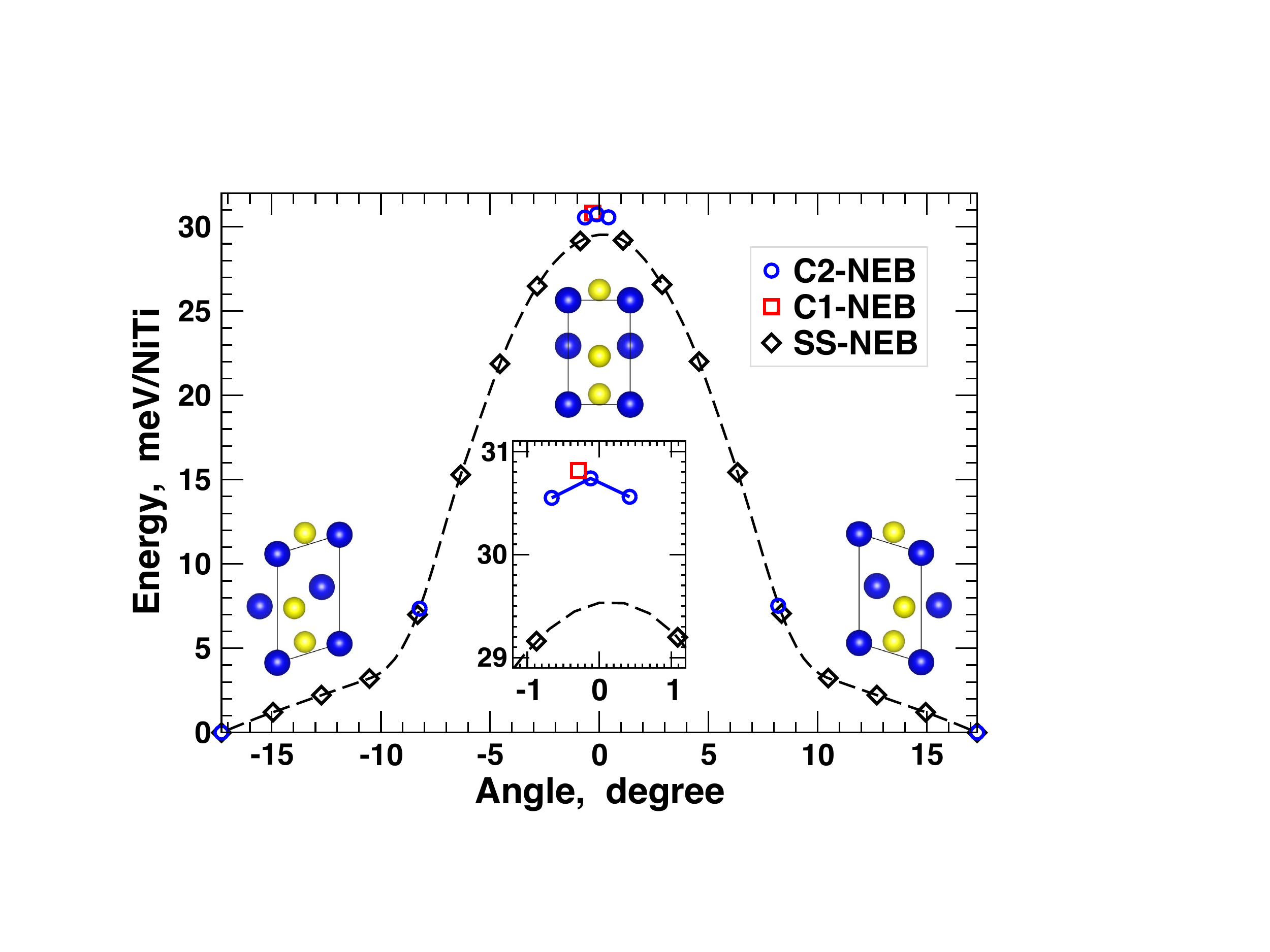}
\caption{\label{Fig3} 
Energy of deformation of the NiTi BCO structure \cite{BCO} in 16-image SS-NEB (black diamonds), 1-image C1-NEB (red square), and 5-image C2-NEB (blue circles) vs. angle between the lattice vector $a$ and the normal to $(bc)$ plane.  
Insets: enlarged near TS; unit cells of TS (B19) and terminal BCO (B19') structures with 
Ni (yellow) and Ti (blue) atoms. Lines (dashed splines and solid connectors) are guide to the eye. }
\end{figure}
\begin{figure}
\includegraphics[width=80mm]{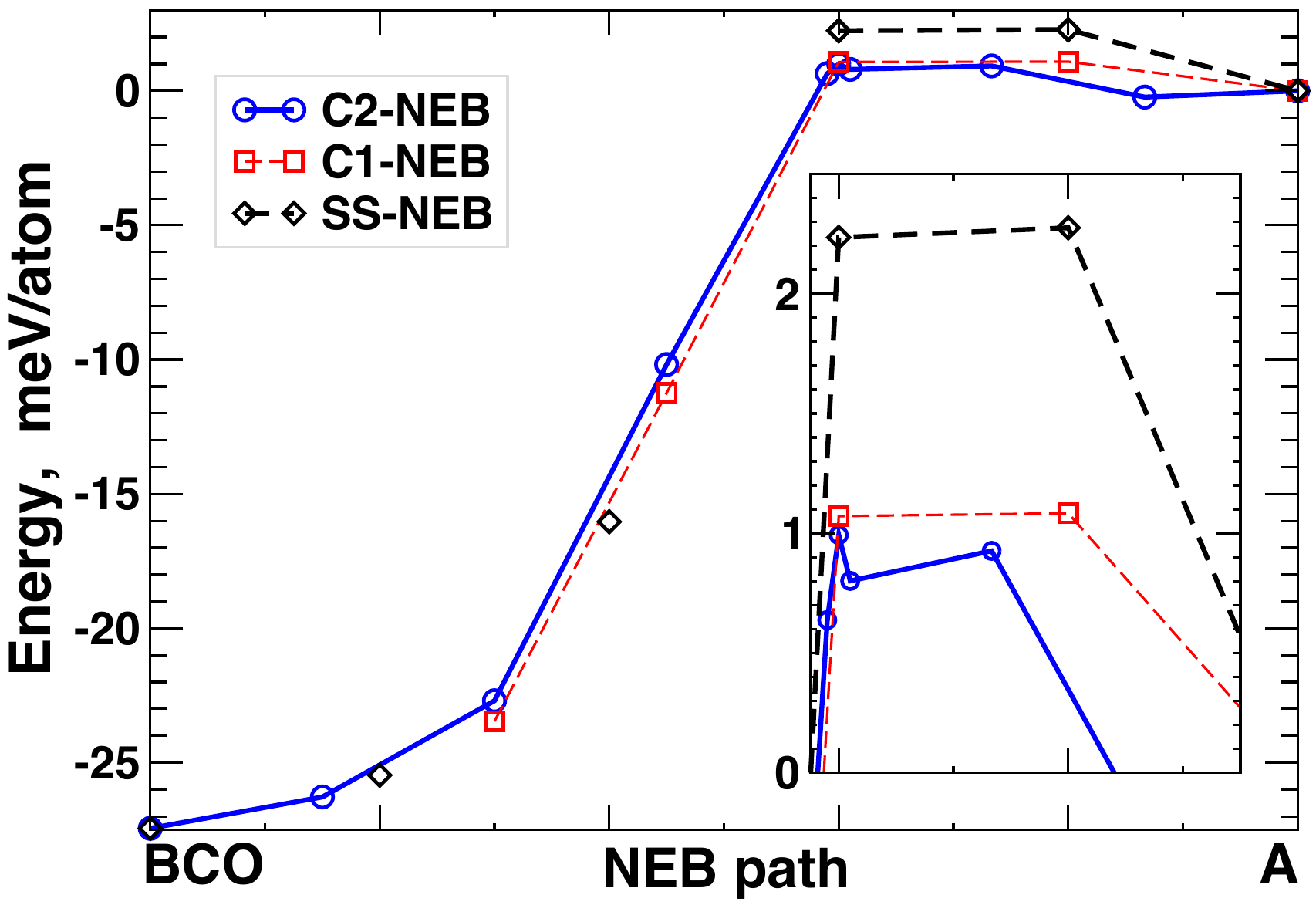}
\caption{\label{Fig4} 
Energy (meV/atom) relative to the austenite (A) for the 
martensitic transformation in NiTi from the low-T BCO \cite{BCO} to the high-T stable austenitic structure \cite{NiTi0} 
in a $108$-atom Ni$_{54}$Ti$_{54}$ unit cell with 324 degrees of freedom.
C2-NEB find the TS at 1 meV/atom. 
C1-NEB swaps the HEI and than diverges; shown is the moment of this swap (not the final C1-NEB result). 
SS-NEB with 4 images overestimates the barrier.  
 }
\end{figure}

\section{Application}
{\par} Below we apply our C2-NEB code to study deformation 
of the martensitic structure in the NiTi shape-memory alloy, \cite{NiTi0,BCO,NiTi1,NiTi2,NAZ2014}  
used as an illustrative example. 
The DFT details are specified in Ref.~\onlinecite{NiTi0}. 
Equiatomic NiTi has a base-centered orthorhombic (BCO) ground state, \cite{BCO} which is easily deformed into the observed B19' structure in the martensite. 
Deformation energies,  obtained from the SS-NEB without climbing images and from the C1-NEB, and C2-NEB algorithms, are shown in Fig.~\ref{Fig3}. All 3 methods give the energy barrier at $30 \pm 1\,$meV/NiTi, and C1-NEB  is as good as C2-NEB in this simple case. 

{\par }  Considering a more complicated structure of the stable NiTi austenite \cite{NiTi0} with hexagonal Ni$_{27}$Ti$_{27}$ primitive cell, we find that SS-NEB overestimates the barrier, while C1-NEB misses the TS for the Austenite-to-Martensite transition (Fig.~\ref{Fig4}). 
Calculations are performed in a $108$-atom Ni$_{54}$Ti$_{54}$ unit cell, constructed from Ni$_{27}$Ti$_{27}$ primitive cell doubled along $a$, which transforms to 27 BCO 4-atom unit cells. 
This cell has 324 degrees of freedom that are coupled, 
making it computationally demanding as the number of NEB images increase, 
but still possible with a limited number of images in C2-NEB.

{\par }  First, we use the standard SS-NEB \cite{GSSNEB} with 4 images (without any climbing images), and get a barrier above 2~meV/atom (Fig.~\ref{Fig4}). 
Next,  we change the algorithm to C1-NEB \cite{C1NEB} with the same number of images, and the single climbing image runs away to unreasonably high energies. 
We discard this diverged unsuccessful result of C1-NEB. 
We take the same result of SS-NEB, \cite{GSSNEB} add more images as linear interpolations between available structures, and 
use C2-NEB with 8 images to find a barrier of 1~meV/atom.  
C2-NEB has 3 images at the TS, including two climbing images and the HEI nudged to them.  
We combine those 3 images into one, and feed the result of C2-NEB into the C1-NEB  code. \cite{C1NEB}
Given this initial configuration, C1-NEB moved images along its NEB path, swapped the HEI, and diverged again:  the moment of this swap is shown in Fig.~4, not the final divergent state. 
Adding or removing a nudged image near the austenite did not stabilize C1-NEB and did not prevent the HEI swaps. 

{\par}In both C2-NEB and in C1-NEB, we find that each change of the HEI index slows down convergence, because all other (nudged) images try to rearrange equidistantly between the new climbing image and a fixed terminal state. 
In C1-NEB, a swap of the climbing HEI can change convergence to divergence.  
However, in C2-NEB we did not observe any attempts of divergent climbing to unreasonably high energies. 
Hence, C2-NEB is more stable. 
With sufficient accuracy at a small computational cost, C2-NEB can be used with a small total number of images (minimum 3; we used 5 or more), while its 3 images approaching the TS create a dense grid at the right place, improving convergence.

\section{Summary}
In summary, we have implemented a two-climbing-images NEB (C2-NEB) algorithm to find reliably minimum-energy pathways in complex potential-energy landscapes. We tested our C2-NEB method on the martensite-to-martensite and austenite-to-martensite transformations in NiTi, a shape-memory alloy, with a serpentine-like minimum-energy pathway for the complex austenite-to-martensite phase transition. 
C2-NEB finds a transition state on a serpentine pathway even if C1-NEB misses it.
If C1-NEB finds the TS, then C2-NEB finds it with higher stability and accuracy.  
C2-NEB is more reliable than C1-NEB because the climbing images move along the minimum-energy pathway and form a dense grid of images near the transition state.  

{\par} The implementation of the C2-NEB algorithm is provided as a useful modification of the existing NEB \cite{C1NEB} and the generalized solid-state NEB (SS-NEB) \cite{GSSNEB} codes. 
The C2-NEB code is available as a replacement of {\tt{neb.F}} file in the NEB code; \cite{C1NEB} as implemented in VASP, the two-climbing-images algorithm is turned on by setting {\tt{LCLIMB=.TRUE.}} in the INCAR file. \cite{VASP}


\section*{Acknowledgments}
This work is supported by the U.S. Department of Energy, Office of Science, Basic Energy Sciences, Materials Science and Engineering Division. The research is performed at the Ames Laboratory, which is operated by Iowa State University under contract DE-AC02-07CH11358.


\end{document}